\documentclass[12pt,tcilatex]{article}

\protect
\input{tcilatex}
\begin{document}
\date{\bf
\scriptsize Rendiconti Lincei - Matematica E Applicationi,\ s. 9, v. 14, pp. 69-83 (2003)}
\title{Hamiltonian Principle\\ in
 Binary Mixtures of Euler Fluids \\
with Applications to the Second Sound Phenomena}
\author{Henri Gouin\,\thanks{henri.gouin@univ-cezanne.fr\qquad\qquad\qquad\qquad\qquad\qquad\qquad\qquad\qquad\qquad\qquad\qquad\qquad\qquad
\qquad\qquad\qquad\qquad\qquad\qquad\qquad\qquad\qquad\qquad\qquad:
On leave from University of Aix - Marseille, CNRS UMR 6181, 13397 Marseille Cedex
20, France.
} \ and Tommaso Ruggeri\,\thanks{ruggeri@ciram.unibo.it\qquad http://www.ciram.unibo.it/ruggeri} \\
{\small {Department of Mathematics and Research Center of Applied
Mathematics (C.I.R.A.M.)} }\\
{\small {University of Bologna, Via Saragozza 8, 40123 Bologna, Italy}}}
\maketitle

\begin{abstract}
In the present paper we compare the theory of mixtures based on
Rational Thermomechanics with the one obtained by
\textsc{Hamilton} principle. We prove that the two theories
coincide in the adiabatic case when the action is constructed with
the \emph{intrinsic Lagrangian}. In the complete thermodynamical
case we show that we have also coincidence in the case of low
temperature when the second sound phenomena arises for superfluid
Helium and crystals.
\end{abstract}

\section{Introduction}

The first mathematical model of homogeneous mixture of fluids in
the context of Rational Thermodynamics was due to
\textsc{Truesdell} \cite{trusd}. The compatibility with the second
principle of thermodynamics was well established by
\textsc{M\"{u}ller} in the framework of classical mechanics
\cite{ingomixture} and by \textsc{Hutter} and \textsc{M\"{u}ller}
in relativity \cite{koli}.

In the framework of  binary mixture of \textsc{Euler} fluids,
\textsc{Dreyer} \cite{dreyer,45} was able to revisit the well
known \textsc{Landau} model of superfluidity
\cite{landau,putterman}. The second sound phenomena in the case of
liquid He II is now well explained from a macroscopic point of
view. Recently \textsc{Ruggeri} \cite{rogers} observed that a
mixture of two \textsc{Euler} fluids can be regarded as a single
heat conducting fluid. This result is advantageous to explain the
second sound phenomena of crystals with the same model than for
superfluid helium.

A different approach was given by \textsc{Gavrilyuk} \emph{et al},
\cite{ggp},  \textsc{Gavrilyuk} and \textsc{Gouin} \cite {gg1,
gg2}. They consider a variational approach to describe
two-velocity effects in homogeneous mixtures: a Lagrangian of the
system is chosen  as a difference of the kinetic energy of the two
constituents and a volumic potential which is Galilean invariant
depending on the relative velocity of components. The equation of
motions of the two components are not in balance form (in fact
they are in balance form in Lagrangian variables associated with
each component). Nevertheless, the  momentum and the energy
equations for the total mixture are in the clasical balance form.

The present work compares the previous approaches and proves that
the two theories coincide in the mechanical case when the
Hamiltonian action is constructed with the \emph{intrinsic
Lagrangian}, i.e. does not depend on the relative velocity. Such
is the case with the Lagrangian considered by \textsc{Gouin} in
\cite{g}. In the thermodynamical case we prove also the
coincidence
 in the case of low temperature and we obtain a complete
agreement between the two approaches and the superfluid model
considered first by \textsc{Landau}.

\section{The Binary Mixtures of Euler Fluids}

The thermodynamics of a homogeneous mixture of $n$ constituents is
well codified as a branch of Extended Thermodynamics \cite{et}. It
is based on the metaphysical principles of \textsc{Truesdell} [1]
 which  postulates the same balance laws of a single fluid for simple mixtures.

\subsection{The Balance System}

The equations of balance of mass, momentum and energy of the
constituents read as follows
 {\normalsize
\begin{eqnarray}
&&\frac{\partial \rho _{a}\ }{\partial t}+\mathrm{div}\left( \rho _{a}
\mathbf{v}_{a}\right) =\tau _{a},  \nonumber \\
&&  \nonumber \\
&&\frac{\partial \rho _{a}\mathbf{v}_{a}}{\partial
t}+\mathrm{div}\left( \rho _{a}\mathbf{v}_{a}\mathbf{\otimes
v}_{a}-\mathbf{t}_{a}\right) =\mathbf{
\ m}_{a},\quad \quad \qquad (a=1,2,\ldots n),  \label{mixture} \\
&&  \nonumber \\
&&\frac{\partial \left( \frac{1}{2}\rho _{a}v_{a}^{2}+\rho _{a}\varepsilon
_{a}\right) }{\partial t}+\mathrm{div}\left\{ \left( \frac{1}{2}\rho
_{a}v_{a}^{2}+\rho _{a\ }\varepsilon _{a}\right) \mathbf{v}_{a}-\mathbf{t}
_{a}\mathbf{v}_{a}+\mathbf{q}_{a}\right\} =e_{a.}  \nonumber
\end{eqnarray}}
\noindent These equations have the same form as  the balance
equations for a single body, except for the non-zero right hand
sides which represent the production  of masses, momenta and
energies. These productions are due to interaction between the
different constituents. Of course, since the total mass, momentum
and energy of the total mixture is conserved, we must have
\[
\sum_{a=1}^{n}\tau _{a}=0,\qquad
\sum_{a=1}^{n}\mathbf{m}_{a}=0,\qquad \sum_{a=1}^{n}e_{a}=0.
\]
where $\rho _{a},\mathbf{v}_{a},\varepsilon
_{a},\mathbf{t}_{a},\mathbf{q}_{a}$ are the mass density,
velocity, internal energy, stress and heat flux respectively of
the  $a$-component of the mixture.

If we sum the equations $(\ref{mixture})$ over all constituents
and introduce
\begin{eqnarray}
&&\hbox{the density}\quad \rho =\sum_{a=1}^{n}\rho _{a},\qquad
\hbox{the velocity}\quad \mathbf{v}=\sum_{a=1}^{n}\frac{\rho
_{a}}{\rho }\ \mathbf{v}
^{a},  \label{macros} \\
&&\hbox{the diffusion velocity}\quad \mathbf{u}_{a}=\mathbf{v}_{a}-\mathbf{v}
,  \label{diffvel} \\
&&\hbox{the stress tensor}\quad \mathbf{t}=\sum_{a=1}^{n}\left( \mathbf{t}
_{a}-\rho _{a}\mathbf{u}_{a}\mathbf{\otimes u}_{a}\right) ,  \label{stress}
\\
&&\hbox{the intrinsic energy density}\;\rho \varepsilon
_{I}=\sum_{a=1}^{n}\rho _{a}\varepsilon _{a}, \\
&&\hbox{the internal energy density }\quad \rho \varepsilon =\rho
\varepsilon _{I}+\frac{1}{2}\sum_{a=1}^{n}\rho _{a}u_{a}^{2},
\label{inten}
\\
&&\hbox{and the heat flux }\quad \mathbf{q}=\sum_{a=1}^{n} \left\{
\mathbf{q} _{a}+\rho _{a}(\varepsilon _{a}+\frac{1}{2}u_{a}^{2})
\mathbf{u}_{a}-\mathbf{t}_{a}\mathbf{u}_{a}\right\} ,
\label{hflux}
\end{eqnarray}
we obtain for the total mixture:\\

The  balance mass
\begin{equation}
\frac{\partial \rho }{\partial t}+\mathrm{div}\left( \rho \mathbf{v}\right)
=0,  \label{mass}
\end{equation}

The balance equation of momentum
\begin{equation}
\frac{\partial \rho \mathbf{v}}{\partial t}+\mathrm{div}\left(
\rho \mathbf{ \ v\otimes v}-\mathbf{t}\right) =0,
\label{momentum}
\end{equation}

The balance of energy
\begin{equation}
\frac{\partial \left( \frac{1}{2}\rho v^{2}+\rho \varepsilon \right) }{
\partial t}+\mathrm{div}\left\{ \left( \frac{1}{2}\rho v^{2}+\rho
\varepsilon \right) \mathbf{v}-\mathbf{tv}+\mathbf{q}\right\} =0.
\label{energy}
\end{equation}
\\
Note that  equations (\ref{mass}, \ref{momentum}, \ref{energy})
 have the same form as those for a single fluid.
Moreover in  equation (\ref{energy}) for the balance of  energy we
observe that the total kinetic energy is $\displaystyle
\frac{1}{2}\rho v^{2}$ is not the sum of the kinetic energy of the
components. In fact we have
\[
\frac{1}{2}\rho v^{2}=\frac{1}{2}\sum_{a=1}^{n}\rho _{a}v_{a}^{2}-\frac{1}{2}
\sum_{a=1}^{n}\rho _{a}u_{a}^{2}.
\]
By analogy with the intrinsic internal energy we call
\emph{intrinsic kinetic energy} the expression
\[
E_c=\frac{1}{2}\sum_{a=1}^{n}\rho _{a}v_{a}^{2}.
\]
As we consider a single absolute temperature $T,$ the aim of
extended thermodynamics for fluid mixtures is the determination of
the $4n+1$ fields :
$$
\begin{array}{ll}
\hbox{mass densities}\qquad & \rho _{a} \\
\hbox{velocities} & \mathbf{v}_{a}\qquad (a=1,2,\ldots n). \\
\hbox{temperature} & T
\end{array}
$$
To determinate these fields we need an appropriate number of
equations. They are based on the equations for each constituent of
balance of mass (\ref {mixture})$_{1}$, momentum
(\ref{mixture})$_{2}$ and conservation of energy of the total
mixture (\ref{energy}).

\subsection{The Equations of Binary Mixture of Euler Fluids}

We consider a binary mixture of \textsc{Euler} fluids, i.e. fluids
that are neither viscous nor heat-conducting :
\[
\mathbf{q}_{a}\equiv 0,\qquad
\mathbf{t}_{a}=-p_{a}\mathbf{I},\qquad (a=1,2).
\]
Instead of the mass and momentum balance laws for the second
component, we use the equivalent equations of total conservation
 for mass and momentum. Therefore, associated with
 the $9$ unknown fields ($\rho _{1},\rho
_{2},\mathbf{v}_{1},\mathbf{v}_{2},T$), we have the  $9$ balance
equations: {\normalsize
\begin{eqnarray}
&&\frac{\partial \rho \ }{\partial t}+\mathrm{div}\left( \rho \mathbf{v}
\right) =0  \nonumber \\
&&  \nonumber \\
&&\frac{\partial \rho_{1}\ }{\partial t}+\mathrm{div}\left( \rho _{1}
\mathbf{v}_{1}\right) =\tau _{1}  \nonumber \\
&&  \nonumber \\
&&\frac{\partial \rho \mathbf{v}}{\partial t}+\mathrm{div}\left( \rho
\mathbf{v\otimes v}-\mathbf{t}\right) =0  \label{table1} \\
&&  \nonumber \\
&&\frac{\partial \rho _{1}\mathbf{v}_{1}}{\partial t}+\mathrm{div}\left(
\rho _{1}\mathbf{v}_{1}\mathbf{\otimes v}_{1}+p_{1}\mathbf{I}\right) =
\mathbf{m}_{1}  \nonumber \\
&&  \nonumber \\
&&\frac{\partial \left( \frac{1}{2}\rho v^{2}+\rho \varepsilon \right) }{
\partial t}+\mathrm{div}\left\{ \left( \frac{1}{2}\rho v^{2}+\rho
\varepsilon \right) \mathbf{v-tv}\ +\mathbf{q}\right\} =0
\nonumber
\end{eqnarray}
} with {\normalsize
\begin{eqnarray}
&&\mathbf{q}=\sum_{a=1}^{2}\left\{ \rho _{a}\left( \varepsilon _{a}+\frac{1}{
2}u_{a}^{2}\right) +p_{a}\right\} \mathbf{u}_{\alpha },\qquad  \nonumber \\
&&\mathbf{t}=-\sum_{a=1}^{2}\left( p_{a}\mathbf{I+}\rho _{a}\mathbf{u}_{a}
\mathbf{\otimes u}_{a}\right) ,  \label{table2} \\
&&p=\sum_{a=1}^{2}p_{\alpha }.  \nonumber
\end{eqnarray}
}

\subsection{The Entropy Principle and Thermodynamical Restrictions}

The compatibility between the system (\ref{mixture}) and the
entropy principle expresses in the form
\begin{equation}
\frac{\partial \rho S}{\partial t}+\mathrm{div}\left\{ \rho
S\mathbf{v}+ \mathbf{\Psi }\right\} \geq 0,  \label{entrop}
\end{equation}
which yields several restrictions on the constitutive equations
\cite{et} :

\begin{eqnarray}
\rho S &=&\rho _{1}S_{1}+\rho _{2}S_{2}  \label{entropic1} \\
&&  \nonumber \\
p_{1} &\equiv &p_{1}(\rho _{1},T);\;p_{2}\equiv p_{2}(\rho
_{2},T);\;\varepsilon _{1}\equiv \varepsilon _{1}(\rho
_{1},T);\;\varepsilon _{2}\equiv \varepsilon _{1}(\rho _{2},T)\;
\label{entropic2}
\end{eqnarray}
such that
\begin{eqnarray}
TdS_{1} &=&d\varepsilon _{1}-\frac{p_{1}}{\rho _{1}^{2}}\ d\rho
_{1};\;\,TdS_{2}=d\varepsilon _{2}-\frac{p_{2}}{\rho _{2}^{2}}\
d\rho _{2}\;
\label{entropic3} \\
&&  \nonumber \\
\mathbf{\Psi } &=&\frac{\mathbf{q}}{T}-\frac{1}{T}\left( \rho
_{1}\mu _{1} \mathbf{u}_{1}+\rho _{2}\mu _{2}\mathbf{u}_{2}\right)
.  \label{entropic4}
\end{eqnarray}
where $\displaystyle \mu _{a}\equiv \varepsilon
_{a}+\frac{p_{a}}{\rho _{a}}-TS_{a}$ is the chemical potential of
constituent $a$.
\subsection{The Mixture considered as a Single Heat conducting Fluid}

\textsc{Ruggeri} \cite{rogers} proved that it is possible to write
the velocities of the two constituents in terms of mass velocity
and heat flux centers :
\[
\mathbf{v}_{1} =\mathbf{v\ +\ }\frac{\alpha }{\rho _{1}}\
\mathbf{q}, \qquad \mathbf{v}_{2} =\mathbf{v\ -\ }\frac{\alpha
}{\rho _{2}}\ \mathbf{q}
\]
where
\begin{equation}
\frac{1}{\alpha} ={\left( \varepsilon _{1}+\dfrac{p_{1}}{\rho _{1}}+\dfrac{1
}{2}u_{1}^{2}\right) -\left( \varepsilon _{2}+\dfrac{p_{2}}{\rho _{2}}+
\dfrac{1}{2}u_{2}^{2}\right) }.  \label{alpha}
\end{equation}
Introducing the concentration $ \displaystyle c=\frac{\rho
_{1}}{\rho }, $  equations \textrm{(\ref{table1})}$_2$ and
\textrm{(\ref{table1})}$_4$ can be written in terms of $\rho
,c,\mathbf{v}$ and $\mathbf{q}$ and  the system (\ref{table1})
becomes:

{\normalsize
\begin{eqnarray}
&&\frac{\partial \rho \ }{\partial t}+\mathrm{div}\left( \rho \mathbf{v}
\right) =0  \nonumber \\
&&  \nonumber \\
&&\frac{\partial (\rho c)}{\partial t}+\mathrm{div}\left( \rho c\mathbf{v+}
\alpha \mathbf{q}\right) =\tau  \nonumber \\
&&  \nonumber \\
&&\frac{\partial \rho \mathbf{v}}{\partial t}+\mathrm{div}\left( \rho
\mathbf{v\otimes v}\ +p\mathbf{I+}\frac{\alpha ^{2}}{\rho c(1-c)}\mathbf{\
q\otimes q}\right) =0  \label{finali} \\
&&  \nonumber \\
&&\frac{\partial (\rho c\mathbf{v+}\alpha \mathbf{q})}{\partial
t}+\mathrm{\ div}\left\{ \rho c\mathbf{v\otimes v+}\frac{\alpha
^{2}}{\rho c}\mathbf{\ q\otimes q+}\alpha \left( \mathbf{v\otimes
q+q\otimes v}\right) +\nu \mathbf{
\ I}\right\} =\mathbf{-}b\mathbf{q}  \nonumber \\
&&  \nonumber \\
&&\frac{\partial \left( \frac{1}{2}\rho v^{2}+\rho \varepsilon \right) }{
\partial t}+\mathrm{div}\left\{ \left( \frac{1}{2}\rho v^{2}+\rho
\varepsilon +p\right) \mathbf{v}+\left( \frac{\alpha ^{2}\mathbf{v\cdot q}}{
\rho c(1-c)}+1\right) \mathbf{q}\right\} =0.  \nonumber \\
&&  \nonumber
\end{eqnarray}}
To eliminate the index $1$, we write as in \cite{rogers},
{\normalsize \ }$\nu =p_{1},\tau =\tau _{1}$ and $ \mathbf{m}_{1}$
$=-b\mathbf{q}$. In an extended thermodynamic model with $9$
fields,
 the binary mixture can be considered as a single heat conducting fluid with a
variable concentration.

Equation of evolution (\ref{finali})$_4$ is a natural extension of
the \textsc{Cattaneo} equation for the heat flux. Thermal inertia
term $\alpha $ together with term $\nu$ have to be interpreted as
new constitutive functions. The advantage of this procedure comes
from the fact that the two functions are now understandable in the
light of mixture theory: term $\nu $ plays the role of
one-component pressure while the thermal inertia term $\alpha $
given in (\ref{alpha}) is the inverse of the difference between
the non-equilibrium enthalpies of the two constituents.

\subsection{The Superfluidity and Second Sound}

\textsc{Dreyer} \cite{dreyer} proved that the \textsc{Landau}
 theory of superfluidity is a particular case of simple
mixtures  with the thermodynamical peculiarities :
\begin{equation}
S_{s}=0;\,\quad \mu _{s}-\mu
_{n}+\frac{1}{2}(\mathbf{v}_{s}-\mathbf{v} _{n})^{2}=0,\quad
\mathbf{m}_{s}=\tau _{s}\mathbf{v}_{s},  \label{vincolient}
\end{equation}
where the indexes $n$ and $s$ correspond to normal and the
superfluid components.\\ By neglecting the quadratic term in the
second equation, in the small diffusion case the two chemical
potential $\mu _{s}$ and $\mu _{n}$ must be equal. Consequently,
the relation $\mu _{s}= \mu _{n}$ allows to obtain one field
variable in terms of the others and it is possible to write
\[
\rho _{s}\equiv \rho _{s}(\rho ,T)
\]
In this case  equation (\ref{table1})$_2$ evaluates the  mass
production value $\tau _{s}$ and the superfluid helium framework
becomes a theory with $8$ fields (i.e. the system is formed by
equations
(\ref{table1})$_1$,(\ref{table1})$_3$,(\ref{table1})$_4$,(\ref{table1})$_5$
or equivalently equations (\ref
{finali})$_{1},$(\ref{finali})$_{3},$(\ref{finali})$_{4},$(\ref{finali})$
_{5}\;$).

The condition (\ref{vincolient})$_{3}$ is the most complex. In
fact (\ref{table1})$_4$ with (\ref{table1})$_2$ can be rewritten
(see \cite{45} for details) :

\[
\frac{\partial \mathbf{v}_{s}}{\partial t}+\mathrm{\nabla }\left(
\frac{1}{2} v_{s}^{2}+\mu _{s}\right) +\text{curl}\ \mathbf{v}_s
\times
 \mathbf{v}_{s}=0.
\]
This equation is in balance form only when the involutive
constraint $ \text{curl}\ \mathbf{v}_{s} =0$ holds. In this case
the system $(\ref{finali})$ coincides with the Landau model
\cite{landau} : {\normalsize
\begin{eqnarray}
&&\frac{\partial \rho \ }{\partial t}+\mathrm{div}\left( \rho
\mathbf{v}
\right) =0,  \nonumber \\
&&  \nonumber \\
&&\frac{\partial \rho \mathbf{v}}{\partial t}+\mathrm{div}\left( \rho
\mathbf{v\otimes v}-\mathbf{t}\right) =0,  \nonumber \\
&&  \label{landau} \\
&&\frac{\partial \mathbf{v}_{s}}{\partial t}+\mathrm{\nabla
}\left( \frac{1}{
2}v_{s}^{2}+\mu _{s}\right) \ =0,  \nonumber \\
&&  \nonumber \\
&&\frac{\partial \left( \frac{1}{2}\rho v^{2}+\rho \varepsilon
\right) }{
\partial t}+\mathrm{div}\left\{ \left( \frac{1}{2}\rho v^{2}+\rho
\varepsilon \right) \mathbf{v}-\mathbf{tv}\ +\mathbf{q}\right\}
=0. \nonumber
\end{eqnarray}}
Taking into account  (\ref{entrop}), (\ref {entropic4}) and
(\ref{vincolient})$_{1}$, the entropy law reduces to the
\textsc{Clausius} form:

\begin{equation}
\frac{\partial \rho S}{\partial t}+\mathrm{div}\ \left( \rho S\
\mathbf{v}+ \frac{\mathbf{q}}{T}\right) =0.  \label{landauentr}
\end{equation}
where  the heat flux (\ref{table2})$_1$ is:

\begin{equation}
\mathbf{q}=\rho TS\ \mathbf{u}_{n}+\frac{1}{2}\left( \rho
_{s}u_{s}^{2}\mathbf{ u}_{s}\mathbf{+}\rho
_{n}u_{n}^{2}\mathbf{u}_{n}\right) . \label{quetto}
\end{equation}
In the diffusion velocity  we neglect the third order terms and we
obtain the {\normalsize \textsc{Landau }} entropy law for the heat
flux \cite{landau}.  The entropy flux becomes  $\rho S\
\mathbf{v}_{n}$ and the entropy is convected by the normal
component
\begin{equation}
\frac{\partial \rho S}{\partial t}+\mathrm{div}\ \left( \rho S\
\mathbf{v}_n \right) =0.  \label{entrop}
\end{equation}
To focus on the thermal wave associated with the second sound we
consider a rigid body  at rest with constant density. For the
superfluid component, the system of energy and momentum equations
is :
\begin{eqnarray*}
&&\frac{\partial \rho \varepsilon }{\partial t}+\mathrm{div}\ \mathbf{q}=0, \\
&& \\
&&\frac{\partial \mathbf{v}_{s}}{\partial t}+\mathrm{\nabla }\left( \frac{1}{
2}v_{s}^{2}+\mu _{s}\right) =0, \\
&&
\end{eqnarray*}
with $ \mathbf{q}=\rho TS\ \mathbf{v}_{n}. $ Such a system is in
the form (\ref{finali}) for a single fluid : {\normalsize
\begin{eqnarray*}
&&\frac{\partial \rho \varepsilon }{\partial t}+\mathrm{div}\ \mathbf{q}=0 \\
&& \\
&&\frac{\partial (\alpha \mathbf{q})}{\partial t}+\mathrm{\nabla }\nu =
\mathbf{-}b\mathbf{q}
\end{eqnarray*}
The system coincides with the one deduced by \textsc{Ruggeri} and
coworkers for the model of second sound in crystals \cite{rugg4}.
Such a model  explains the change of form of the initial square
thermal waves both in crystals \cite{rugg4, rugg3, rugg5} and in
the superfluid helium \cite{rugg7}.}

\section{The Hamiltonian Procedure for Two-Fluid Mixtures
}

To obtain the equations of motion and energy, the procedure is the
following:\\ Let us suppose that the mixture of two miscible
fluids is well described by the two-component velocities
$\mathbf{v}_{1}, \mathbf{v} _{2}$, the densities $\rho _{1}, \rho
_{2}$ and the \textit{
intrinsic internal energy} $\beta =\rho \varepsilon _{I}$. \\
The intrinsic internal energy is a Galilean invariant and does not
depend on the reference frame. We consider the general case where
 $\beta$ depends on $\rho _{1},\rho _{2}$
but also of the relative velocity
$\mathbf{w}=\mathbf{v}_{1}-\mathbf{v}_{2}$ through the norm $
\omega =|\mathbf{v}_{1}-\mathbf{v}_{2}|$ \cite{ggp}. The
\textit{intrinsic kinetic energy}  is $\displaystyle
E_{c}={\frac{1}{2}}\left( \rho _{1}v_{1}^{2}+\rho
_{2}v_{2}^{2}\right) $. \\
 Without dissipative
effects, chemical reactions and with conservation of masses of the
two components, an extended form of \textsc{Hamilton} principle of
least action is used in the form
\[
\delta I=0\quad \mathrm{with}\quad I=\int_{\mathcal{W}_0 }L\,dxdt,
\]
where the Lagrangian is $ L=E_{c}-\beta (\rho _{1},\rho
_{2},\omega )$, \ $\mathcal{W}=[t_{0},t_{1}]\times D\ $ is a
time-space cylinder and the variations must vanish on the boundary
of $\mathcal{W}$. The virtual motions of the mixture are defined
in \cite{ggp,gg1}.\\ From the variations of \textsc{Hamilton}
action, we obtain the equations of motions in the form

\begin{equation}
{\frac{\partial \mathbf{k}_{a}}{\partial t}}+\mathrm{curl}\
\mathbf{k}_{a}\times \mathbf{v}_{a}+\nabla \left( {\frac{\partial
\beta}{\partial \rho _{a} }}-{\frac{1}{2}}\
v_{a}^{2}+\mathbf{k}_{a} \mathbf{v}_{a}\right) =0\;\;\quad \quad
(a=1,2)  \label{(2)}
\end{equation}
where
\[
\mathbf{k}_{a}=\mathbf{v}_{a}-(-1)^{a}{\frac{1}{\rho
_{a}}}{\frac{\partial \beta }{\partial \omega
}}{\frac{\mathbf{w}}{\omega }}.
\]
The momentum conservation law is obtained by summing on $a=1,2$
equation (\ref{(2)}) multiplied by $\rho _{a}$ : {\normalsize
\begin{eqnarray}
{\frac{\partial \left( \rho _{1}\mathbf{v}_{1}+\rho _{2}\mathbf{v}
_{2}\right) }{\partial t}}& + &  \; \nabla \left( \rho
_{1}{\frac{\partial \beta}{\partial \rho _{1}}}+\rho
_{2}{\frac{\partial \beta}{\partial \rho
_{2}}}-\beta\right)   \nonumber\\
&+&\text{\textrm{div}}\left( \rho _{1}\mathbf{v} _{1}\otimes
\mathbf{v}_{1}+\rho _{2}\mathbf{v}_{2}\otimes \mathbf{v}_{2}-{
\frac{\partial \beta}{\partial \omega }}{\frac{{\mathbf{w}\otimes
\mathbf{w}}}{ \omega }}\right)=0 \label{(3)}
\end{eqnarray}
Additive terms  come from the dependance of $\beta$ in $\omega $
and in the mechanical case $\displaystyle \rho _{1}{\frac{\partial
\beta}{\partial \rho _{1}}}+\rho _{2} {\frac{\partial
\beta}{\partial \rho
_{2}}}- \beta$ represents the total pressure $p$. \\
The  conservation of energy  is obtained by summing on $a=1,2$
equation (\ref{(3)}) multiplied by $\rho _{a}\mathbf{v}_{a}$ :
\begin{equation}
\frac{{\partial }}{\partial t}{\left( {\frac{1}{2}}\rho
_{1}{v}_{1}^{2}+{ \frac{1}{2}}\rho _{2}{v}_{1}^{2}+\beta+\omega
{\dfrac{\partial \beta}{\partial \omega }} \right)
}+\mathrm{div}\left( \rho _{1}\mathbf{v}_{1}{\frac{\partial \beta
}{\partial \rho _{1}}}+\mathbf{k}_{1} \mathbf{v}_{1}+\rho
_{2}\mathbf{v} _{2}{\frac{\partial W}{\partial \rho
_{2}}}+\mathbf{k}_{2} \mathbf{v} _{2}\right) =0.  \label{(4)}
\end{equation}
In paragraph 2, we consider the case where $\beta$ is independent
of $\omega $ and the entropy principle (\ref{entropic2}) presented
in \cite{et} yields $\beta =\rho _{1}\varepsilon _{1}(\rho
_{1})+\rho _{2}\varepsilon _{2}(\rho _{2})$. Then, equation
(\ref{(2)}) writes
\begin{equation}
{\frac{\partial \mathbf{v}_{a}}{\partial t}}+\mathrm{curl}\
\mathbf{v} _{a}\times \mathbf{v}_{a}+\nabla \left( {\frac{1}{2}}\
v_{a}^{2}+\mu _{a}\right) =0,\quad \quad (a=1,2).  \label{(5)}
\end{equation}
Multiplying equation (\ref{(5)}) by $\rho _{a}$ straightforward
calculations yield equation (\ref{table1})$_4$ with
$\mathbf{m}_{a}=0$. Equations (\ref{(3)}, \ref{(4)}) yield
equations (\ref{table1})$_3$, (\ref{table1})$_5$ and balance of
mass equations correspond to  $ \tau _{a}=0\; \; (a =1,2)$.\\
A purely mechanical case is the adiabatic one and we have verified
the following results :\\
\emph{In the adiabatic case with intrinsic Lagrangian} $
L=E_{c}-\rho \varepsilon _{I} $ \emph{difference between the
intrinsic kinetic energy and the intrinsic internal energy} $\rho
\varepsilon _{I}=\rho _{1}\varepsilon _{1}(\rho _{1})+\rho
_{2}\varepsilon _{2}(\rho _{2}),$\emph{ \  the system deduced from
Hamilton principle coincides with the system coming from Rational
Thermomechanics.}

\section{The Hamiltonian Procedure for Superfluid Helium}

In the case of a binary mixture some change must be done in the
definition of \emph{virtual motions} presented by \textsc{Serrin}
in  \cite{serrin}. Let us consider the motion of Helium II as two
diffeomorphisms
\[
\mathbf{z}=M\left( \mathbf{Z}\right) ,\quad \mathbf{z}=M_{n}\left(
\mathbf{Z} _{n}\right)
\]

\noindent where \ $\displaystyle \mathbf{z}=\pmatrix{\displaystyle
t\cr \mathbf{x}}$ corresponds to the Eulerian variables in
time-space and  $ \mathbf{Z}=\pmatrix{\displaystyle \lambda \cr
{\bf X}}, \mathbf{Z}_{n}=\pmatrix{\displaystyle \lambda_n \cr {\bf
X}_n}\ $ correspond to the Lagrangian variables associated with
the barycentric motion and the normal component motion of helium
II. In coordinate form,
\[
M\left( \mathbf{Z}\right) =\pmatrix{\displaystyle g\left( \lambda
,\mathbf{X} \right) \cr {\mbox{\boldmath$\phi$}}\left( \lambda
,\mathbf{X}\right) } ,\quad M_{n}\left( \mathbf{Z}_{n}\right)
=\pmatrix{\displaystyle g\left( \lambda _{n},\mathbf{X}_{n}\right)
\cr {\mbox{\boldmath$\phi$}}_{n}\left( \lambda
_{n},\mathbf{X}_{n}\right) }
\]
We consider three  \emph{one-parameter families of virtual
motions} which are sufficient to obtain the governing equations :
$$({\cal F})\
\left\{ \matrix{t=g\left( \lambda ,\mathbf{X}\right) =g_{n}\left(
\lambda _{n},\mathbf{X}_{n}\right) \cr
\mathbf{x}={\mbox{\boldmath$\Phi$}}\left( \lambda
,\mathbf{X},\varepsilon \right) \hfill \cr
\mathbf{x}={\mbox{\boldmath$\phi$}}_{n}\left( \lambda
_{n},\mathbf{X} _{n}\right) \hfill \cr }\right.
$$
with ${\mbox{\boldmath$\Phi$}}\left( \lambda ,\mathbf{X},0\right)
={ \mbox{\boldmath$\phi$}}\left( \lambda ,\mathbf{X}\right) $,
$$ ({\cal F}_n)\
\left\{ \matrix{t=g\left( \lambda ,\mathbf{X}\right) =g_{n}\left(
\lambda _{n},\mathbf{X}_{n}\right) \cr
\mathbf{x}={\mbox{\boldmath$\phi$}}\left( \lambda
,\mathbf{X}\right) \hfill \cr
\mathbf{x}={\mbox{\boldmath$\Phi$}}_{n}\left( \lambda
_{n},\mathbf{X} _{n},\varepsilon \right) \hfill \cr }\right.
$$
with ${\mbox{\boldmath$\Phi$}}_{n}\left( \lambda _{n},\mathbf{X}
_{n},0\right) ={\mbox{\boldmath$\phi$}}_{n}\left( \lambda
_{n},\mathbf{X} _{n}\right) ,$
$$\quad \quad \quad (\mathcal{F}_{t})\
\left\{ \matrix{t=G\left( \lambda ,\mathbf{X},\varepsilon \right)
=G_{n}\left( \lambda _{n},\mathbf{X}_{n},\varepsilon \right) \cr
\mathbf{x}={\mbox{\boldmath$\phi$}}\left( \lambda
,\mathbf{X}\right) \hfill \cr
\mathbf{x}={\mbox{\boldmath$\phi$}}_{n}\left( \lambda
_{n},\mathbf{X} _{n}\right) \hfill \cr }\right.
$$
with $G\left( \lambda ,\mathbf{X},0\right) \ =G_{n}\left( \lambda
_{n}, \mathbf{X}_{n},0\right) =g\left( \lambda ,\mathbf{X}\right)
=g_{n}\left( \lambda _{n},\mathbf{X}_{n}\right) $.
\bigskip

\noindent The three families generate the virtual displacements
\[
{\mbox{\boldmath$\zeta $}}=\pmatrix{\displaystyle 0\cr\cr
{\mbox{\boldmath$\xi$}}}=\left. \left( \matrix{\displaystyle
\displaystyle 0\cr\cr \displaystyle {\frac{\partial
{\mbox{\boldmath$\Phi$}} }{\partial \varepsilon }}}\right) \right|
_{\varepsilon =0},\quad { \mbox{\boldmath$\zeta $}}_{n}=\left(
\matrix{\displaystyle 0\cr\cr {\mbox{\boldmath$\xi$}}_{n}}\right)
=\left. \left( \matrix{\displaystyle 0\cr \cr \displaystyle
{\frac{\partial {\mbox{\boldmath$\Phi$}}_{n}}{\partial \varepsilon
}}} \right) \right| _{\varepsilon =0},\quad {\mbox{\boldmath$\zeta
$}}_{t}=\pmatrix{\displaystyle \tau \cr\cr \mathbf{0}}=\left.
\left( \matrix{\displaystyle {\frac{\partial G}{\partial
\varepsilon }}\cr\cr \mathbf{0}}\right) \right| _{\varepsilon =0}.
\]
The virtual motion  $\left( \mathcal{F}\right)  $ generates an
associated displacement $\delta \mathbf{Z}_{n}$ of the normal
component. Indeed, the relations
\[
\displaystyle  g\left( \lambda ,\mathbf{X}\right) =g_{n}\left(
\lambda _{n}, \mathbf{X}_{1}\right)
\]
\[
{\mbox{\boldmath$\phi$}}_{n}\left( \lambda ,\mathbf{X}_{n}\right)
={ \mbox{\boldmath$\Phi$}}\left( \lambda ,\mathbf{X},\varepsilon
\right)
\]
imply
\[
{\mbox{\boldmath$\zeta$}}=\pmatrix {\ \displaystyle
{\frac{\partial g_{n}}{\partial \lambda _{1}}}\; , \displaystyle
{\frac{\partial g_{n}}{{\partial \mathbf{X}_{n}}}}\cr\cr
\displaystyle {\frac{\partial {
\mbox{\boldmath$\phi$}}_{n}}{\partial {\lambda _{n}}}}\; ,
\displaystyle {\frac{
\partial {\mbox{\boldmath$
\phi$}}_{n}}{\partial \mathbf{X}_{n}}}\cr\cr}\delta \mathbf{Z}_{n}
\]
By using the definition of the deformation gradient $\mathbf{F}$
proposed in Appendix we get
\begin{equation}
\delta \mathbf{Z}_{n}=\mathbf{C}_{n}{\mbox{\boldmath$\zeta
$}}\hskip 1cm \hbox{with}\hskip
1cm\mathbf{C}_{n}=\pmatrix{\displaystyle 0& ,& \mathbf{0}\cr\cr{\
-{\bf F}_n^{-1} \mathbf{ V}_n}& ,& \mathbf{F}_{n}^{-1}} \label{cn}
 \end{equation}
In the same way, virtual motion  $\left( \mathcal{F}_{n}\right) $
generates an associated displacement $\delta _{n}\mathbf{Z}$ of
the barycentric motion
\[
\delta _{n}\mathbf{Z}=\mathbf{C}\ {\mbox{\boldmath$\zeta
$}}_{n}\hskip 1cm \hbox{with}\hskip
1cm\mathbf{C}=\pmatrix{\displaystyle 0& ,& \mathbf{0}\cr{-
\mathbf{F}^{-1} \mathbf{V}}& , &\mathbf{F}^{-1}}
\]
Now,  $H\left( \mathbf{Z},\varepsilon \right)  $ notes a
perturbation of $h\left( \mathbf{Z}\right) $, the variation of $h$
is
\[
\delta h=\left. {\frac{\partial H}{\partial \varepsilon }}\right|
_{\varepsilon =0}
\]
We can also introduce Lagrangian variations corresponding to the
families $ (\mathcal{F}_{n})$ and  $(\mathcal{F}_{t}) :$
\[
\delta _{n}h_{n}=\left. {\frac{\partial H_{n}}{\partial
\varepsilon }} \right| _{\varepsilon =0}\hskip 1cm\hbox{and}\hskip
1cm\delta _{t}h_{t}=\left. {\frac{\partial H_{t}}{\partial
\varepsilon }}\right| _{\varepsilon =0}
\]
The variations of the entropy $S$ is a main step of our model: we
make the physical assumption that the entropy $S$ is defined on
the $\mathbf{Z_{n}}$-space. This result corresponds to equation
(\ref{entrop}) proposed by \textsc{Landau}. Consequently, we
deduce $ \delta _{n}S=0\;  \mathrm{and}\; \delta _{t}S=0 $. From
relation (\ref{cn}) we obtain \[ \delta S={\frac{\partial
S}{\partial \mathbf{Z}_n}}\ \delta \mathbf{Z}_n ={\frac{
\partial S}{\partial \mathbf{x}}}\ {\mbox{\boldmath$\xi$}}
\]
Following the Hamiltonian procedure presented in paragraph 3, we
consider the Lagrangian $L$ as a function of $\rho
,\mathbf{v},\rho _{n},\mathbf{v}_{n},S$ \ ($L=L(\rho
,\mathbf{v},\rho _{n},\mathbf{v}_{n},S)$). Such is the case for
the intrinsic Lagrangian $\displaystyle L={\frac{1}{2}} (\rho
_{n}v_{n}^{2}+\rho _{s}v_{s}^{2})-\beta (\rho ,\rho _{n},S)$ where
$\rho_s$ and $\mathbf{v}_s$ are given by the relations :
\begin{equation}
\rho_s = \rho - \rho_n \hskip 1cm\hbox{and}\hskip 1cm \mathbf{v}_s
= {\frac{\rho \mathbf{v}- \rho_n \mathbf{v}_n}{\rho-\rho_n}}.
\label{bary}
\end{equation}
Consequently,
\[
{\frac{{\partial \mathbf{v}_{s}}}{{\partial \rho
}}}={\frac{1}{\rho _{s}}}( \mathbf{v}-\mathbf{v}_{s}),\quad
{\frac{{\partial \mathbf{v}_{s}}}{{\partial \rho
_{n}}}}={\frac{1}{\rho _{s}}}(\mathbf{v}_{s}-\mathbf{v}_{n}),\quad
{ \frac{{\partial \mathbf{v}_{s}}}{{\partial
\mathbf{v}}}}={\frac{\rho }{\rho _{s}}}\ \mathbf{I},\quad
{\frac{{\partial \mathbf{v}_{s}}}{{\partial \mathbf{
v}_{n}}}}=-{\frac{\rho _{n}}{\rho _{s}}}\ \mathbf{I}
\]
The variation of the Hamilton action corresponding to the first
family is :
\[
\delta I=\int_{{\mathcal{W}}_{0}}\delta \left( L\ \mathrm{det}\
\mathbf{B} \right) dw_{0}
\]
where \ $\displaystyle \mathbf{B}={\frac{\partial
\mathbf{z}}{\partial \mathbf{Z}}}\ $ is the Jacobian of $M$ and \
$\mathcal{W}_{0}\ $ is the associated Lagrangian domain in the\
$\displaystyle \pmatrix{\displaystyle \lambda \cr
\mathbf{X}}$-space. Consequently,
\[
\delta I=\int_{\mathcal{W}_{0}}\left( \delta L+L\ \mathrm{Div}\ {
\mbox{\boldmath$\zeta $}}\right) \mathrm{det}\mathbf{B}\ dw_{0}.
\]
Variations of $L$ come from
\[
\delta L={\frac{\partial L}{\partial \mathbf{v}}}\delta
\mathbf{v}+{\frac{
\partial L}{\partial \mathbf{v}_{n}}}\delta \mathbf{v}_{n}+{\frac{\partial L
}{\partial \rho }}\delta \rho +{\frac{\partial L}{\partial \rho
_{n}}}\delta \rho _{n}+{\frac{\partial L}{\partial S}}\delta S.
\]
with,
\[
{\frac{\partial L}{\partial \mathbf{v}}}=\rho \mathbf{v}_{s},\ \ \
{\frac{
\partial L}{\partial \mathbf{v}_{n}}}=\rho _{n}(\mathbf{v}_{n}-\mathbf{v}
_{s})
\]
\begin{equation}
R={\frac{\partial L}{\partial {\rho
}}}=-{\frac{1}{2}}v_{s}^{2}+\mathbf{v} _{s} \mathbf{v}-\beta_{\rho
_{s}}^{\prime }(\rho _{n},\rho _{s},S), \label{(7)}
\end{equation}
\[
R_{n}={\frac{\partial L}{\partial {\rho
_{n}}}}={\frac{1}{2}}v^{2}+{\frac{1}{ 2}}v_{s}^{2}-\mathbf{v}_{s}
\mathbf{v}_{n}-\beta_{\rho _{n}}^{\prime }(\rho _{n},\rho
_{s},S)+\beta_{\rho _{s}}^{\prime }(\rho _{n},\rho _{s},S),
\]
\[
\rho   T=-{\frac{\partial L}{\partial S}}
\]
Moreover we have,
\[
\delta \mathbf{v}_{n}={\frac{\partial \mathbf{v}_{n}}{\partial
\mathbf{Z}_{n} }}\ \mathbf{C}_{n}\ {\mbox{\boldmath$\zeta
$}}={\frac{\partial \mathbf{v}_{n} }{\partial \mathbf{x}}}\
{\mbox{\boldmath$\xi $}}\quad \mathrm{and}\quad \delta \rho
_{n}={\frac{\partial \rho _{n}}{\partial \mathbf{Z}_{n}}}\
\mathbf{C}_{n}\ {\mbox{\boldmath$\zeta $}}={\frac{\partial \rho
_{n}}{
\partial \mathbf{x}}}\ {\mbox{\boldmath$\xi $}}
\]
Since \ $\displaystyle {\mbox{\boldmath$\zeta $}}=\left(
\matrix{\displaystyle 0\cr {\mbox{\boldmath$\xi$}}}\right) ,\ $ we
get (see Appendix for the variations $\delta \rho $ and $\delta
\mathbf{v} $ variations),
\[
\delta L+L\ \hbox{Div }{\mbox{\boldmath$\zeta $}}={\frac{\partial
L}{\partial \mathbf{v}}}{\frac{d{\mbox{\boldmath$\xi$}}}{
dt}}+{\frac{\partial L}{\partial \mathbf{v}_{n}}}\ {\frac{\partial
\mathbf{v} _{n}}{\partial \mathbf{x}}}\
{\mbox{\boldmath$\xi$}}-\rho \ {\frac{\partial L}{
\partial \rho }}\hbox{ div }{\mbox{\boldmath$\xi$}}\
\]
\[
+{\frac{\partial L}{\partial \rho _{n}}}{\frac{\partial \rho
_{n}}{\partial \mathbf{x}}}\ {\mbox{\boldmath$\xi$}}+L\hbox{ div
}{\mbox{\boldmath$\xi$}}+{\frac{\partial L}{\partial S}}\
{\frac{\partial S}{
\partial \mathbf{x}}}\ {\mbox{\boldmath$\xi$}}\
\]
\[
=\rho \mathbf{v}_{s}\ {\frac{d{\mbox{\boldmath$\xi$}}}{dt}}+\rho
_{n}( \mathbf{v}_{n}-\mathbf{v}_{s}){\frac{\partial
\mathbf{v}_{n}}{\partial \mathbf{x}}}{\
\mbox{\boldmath$\xi$}}-\rho  R\hbox{ div }{
\mbox{\boldmath$\xi$}}+R_{n}{\frac{\partial \rho _{n}}{\partial
\mathbf{x}}}\ { \mbox{\boldmath$\xi$}}+L\hbox{ div
}{\mbox{\boldmath$\xi$}}+{\frac{\partial L}{\partial S}}\
{\frac{\partial S}{\partial \mathbf{x}}}\ {
\mbox{\boldmath$\xi$}}\
\]
By using the expression
\[
\rho
\mathbf{v}_{s}{\frac{d{\mbox{\boldmath$\xi$}}}{dt}}={\frac{\partial
}{
\partial t}}\left( \rho \mathbf{v}_{s}\ {\mbox{\boldmath$\xi$}}\right)
-{ \frac{\partial }{\partial t}}\left( \rho \mathbf{v}_{s} \right)
{ \mbox{\boldmath$\xi$}}+\hbox{ div}\Big( \rho ( \mathbf{v}\otimes
\mathbf{v} _{s} )\ {\mbox{\boldmath$\xi$}}\Big) -\hbox{ div}\left(
\rho \mathbf{v} \otimes \mathbf{v}_{s}\right) \
{\mbox{\boldmath$\xi$}}
\]
we get
\[
\delta L+L\hbox{ div }{\mbox{\boldmath$\xi$}}=
\]
\[
{\frac{\partial }{\partial t}}\left( \rho \mathbf{v}_{s}\
{\mbox{\boldmath$\xi$}}\right) -{\frac{\partial }{\partial
t}}\left( \rho \mathbf{v}_{s}\right) \
{\mbox{\boldmath$\xi$}}+\hbox{ div}\Big( \rho (\mathbf{v}\otimes
\mathbf{v}_{s})\ {\mbox{\boldmath$\xi$}} \Big) -\hbox{ div}\left(
\rho \mathbf{v}\otimes \mathbf{v}_{s}\right) \ {
\mbox{\boldmath$\xi$}}\
\]
\[
+\ \rho _{n}\mathbf{v}_{n}\ {\frac{\partial \mathbf{v}_{n}}
{\partial \mathbf{x} }}\ {\mbox{\boldmath$\xi$}}-\hbox{ div}\left(
\rho R\ {\mbox{\boldmath$\xi$}} \right) +\nabla \left( \rho
R\right) {\mbox{\boldmath$\xi$}}+R_{n}{\frac{
\partial \rho _{n}}{\partial \mathbf{x}}}\ {\mbox{\boldmath$\xi$}}+
\hbox{ div}\left( L{\mbox{\boldmath$\xi$}}\right) \
+{\frac{\partial L}{\partial S} }\ \nabla S\
{\mbox{\boldmath$\xi$}}
\]
\[
-\left( {\frac{\partial L}{\partial \rho }}\nabla \rho
+{\frac{\partial L}{
\partial \mathbf{v}}}{\frac{\partial \mathbf{v}}{\partial \mathbf{x}}}+{
\frac{\partial L}{\partial \rho _{n}}}\nabla \rho
_{n}+{\frac{\partial L}{
\partial \mathbf{v}_{n}}}{\frac{\partial \mathbf{v}_{n}}{\partial \mathbf{x}}
}+{\frac{\partial L}{\partial S}}\nabla S\right)
{\mbox{\boldmath$\xi$}}
\]
and from equations (\ref{(7)}),
\[
\delta L+L\ \hbox{Div }{\mbox{\boldmath$\zeta$ }}=
\]
\[
\left( -{\frac{\partial }{\partial t}}(\rho \mathbf{v}_{s})-\hbox{
div } \left( \rho \mathbf{v}\otimes \mathbf{v}_{s}\right) +\nabla
\left( \rho R\right) -R\nabla \rho -\rho \left({\frac{\partial
\mathbf{v}}{\partial \mathbf{x} }}\right)^*\mathbf{v}_{s}\right)
{\mbox{\boldmath$\xi$}}\
\]
\[
+{\frac{\partial }{\partial t}}\left( \rho \mathbf{v}_{s}{
\mbox{\boldmath$\xi$}}\right) +\hbox{ div}\Big( \rho(
\mathbf{v}\otimes \mathbf{v}_{s})\  {\mbox{\boldmath$\xi$}}\Big)
-\hbox{ div }\left( \rho \ R\ { \mbox{\boldmath$\xi$}}\right)
+\hbox{ div }\left( L\ {\mbox{\boldmath$\xi$}} \right) ,
\]
where $^*$ notes the transposition. Consequently, the first
equation of momentum is
\begin{equation}
{\frac{\partial \mathbf{v}_{s}}{\partial t}}+ \nabla \left(
{\frac{1}{2}}v_{s}^{2}+\beta_{\rho _{s}}^{\prime }\right) =
\mathbf{v}_{s}\times \mathrm{curl}\ \mathbf{v}_s  \label{irrot1}
\end{equation}
If we note $\mu _{s}=\beta_{\rho _{s}}^{\prime }$, when
$\mathbf{v}\approx 0$ , equation (\ref{irrot1}) yields
\begin{equation}
{\frac{\partial \mathbf{v}_{s}}{\partial t}}+ \nabla \left(
{\frac{1}{2}}v_{s}^{2}+\mu _{s}\right) =0 \label{irrot2}
\end{equation}
which is the \textsc{Landau} equation for the superfluid
component. In fact \textsc{Landau} pointed out that Helium II lose
its superfluidity when the velocity is not small enough and the
supplementary term $\mathrm{curl}\ \mathbf{v}_{s}\times \mathbf{v}
\approx 0 $ corresponds to this experimental evidence.

\bigskip

Variations of the Hamilton action are closely the same for the
second family. The variation of the entropy is $\delta _{n}S=0$
and consequently an entropy term is now appearing in the equations
of motion. The second equation of momentum is
\[
\frac{{\partial }}{\partial t}{\left( \rho
_{n}(\mathbf{v}_{n}-\mathbf{v} _{s})\right) }+\hbox{ div }\left(
\rho _{n}\mathbf{v}_{n}\otimes (\mathbf{v}_{n}-\mathbf{v}
_{s})\right)
\]
\begin{equation}
+\ \rho _{n}\Big(\frac{\partial \mathbf{u}_{n}}{{\partial
\mathbf{x}}}\Big)^*(\mathbf{v} _{n}-\mathbf{v}_{s})-\rho
_{n}\nabla R_{n}-\rho \ T\ \nabla S=0 \label{momentum2}
\end{equation}
By summing equations (\ref{irrot1}) and (\ref{momentum2}),
equation (\ref{momentum2}) can be replaced by the balance of total
momentum:
\[
{\frac{\partial }{\partial t}}\Big( \rho \mathbf{v}_{s}+\rho
_{n}(\mathbf{v} _{n}-\mathbf{v}_{s})\Big) + \hbox { div}\left(
\rho \mathbf{v}\otimes \mathbf{v}_{s}+\rho _{s}\mathbf{v}
_{n}\otimes (\mathbf{v}_{n}-\mathbf{v}_{s})-\rho {\frac{\partial
L}{\partial \rho }}-\rho _{n}{\frac{\partial L}{\partial \rho
_{n}}}+L\right) =0.
\]
 Straightforward
calculations yield the equation of momentum
\begin{equation}
{\frac{{\partial \rho \mathbf{v}}}{\partial t}}+\hbox{ div }\left(
\rho \mathbf{v}_{n}\otimes \mathbf{v}_{n}+\rho
\mathbf{v}_{s}\otimes \mathbf{v} _{s}+p\right) =0,
\label{momentum3}
\end{equation}
where $p=\rho _{s}\mu _{s}+\rho _{n}\mu _{n}-\beta$ is the total
pressure, with $\mu _{n}=\beta_{\rho _{n}}^{\prime }$.

\bigskip

Finally, the third family is associated with the vector
displacement \ $\displaystyle {\mbox{\boldmath$\zeta
$}}_{t}=\left( \matrix{\displaystyle \tau \cr 0}\right) .\ $ The
variations of basic variables are calculated in Appendix :
\[
\delta _{t}\mathbf{v}=-\mathbf{v}{\frac{d\tau }{dt}},\quad \delta
_{t}\rho =\rho \ \nabla \tau \ \mathbf{v} ,\quad \delta
_{t}\mathbf{v}_{n}=-\mathbf{v}_{n}{\frac{d_{n}\tau }{dt}},\quad
\delta _{t}\rho _{n}=\rho _{n}\nabla \tau \ \mathbf{v}_{n},\quad
\delta _{t}S=0.
\]
The variation of the \textsc{Hamilton} action is
\[
\delta _{t}I =\int_{\mathcal{W}_{0}}\left( \delta
_{t}L+L{\frac{\partial \tau }{\partial t}}\right) \hbox{ det }B\
dw_{0}
\]
with
\[
\delta _{t}L={\frac{\partial L}{\partial \mathbf{v}}}\delta
_{t}\mathbf{v}+{ \frac{\partial L}{\partial \mathbf{v}_{n}}}\delta
_{t}\mathbf{v}_{n}+{\frac{
\partial L}{\partial \rho }}\delta _{t}\rho +{\frac{\partial L}{\partial
\rho _{n}}}\delta _{t}\rho _{n}+{\frac{\partial L}{\partial
s}}\delta _{t}s
\]
Hence,
\[
\delta _{t}L+L{\frac{\partial \tau }{\partial t}}=-\rho
\mathbf{v}_{s} \mathbf{v}\left( {\frac{\partial \tau }{\partial
t}}+\nabla \tau\ \mathbf{v}\right) -\rho
_{n}(\mathbf{v}_{n}-\mathbf{v}_{s})\mathbf{v} _{n}\left(
{\frac{\partial \tau }{\partial t}}+\nabla \tau\  \mathbf{v}
_{n}\right)
\]
\[
+\rho R\nabla \tau\  \mathbf{v}+\rho _{n}R_{n}\nabla \tau\
\mathbf{v }_{n}+{\frac{\partial }{\partial t}}\left( L \tau
\right) -{\frac{\partial L }{\partial t}}\tau
\]
\[
=-{\frac{\partial }{\partial t}}\left( \rho
\mathbf{v}_{s}\mathbf{v}\ \tau \right) +{\frac{\partial }{\partial
t}}\left( \rho \mathbf{v}_{s}\mathbf{v} \right) \tau -\hbox{
div}\Big( \rho (\mathbf{v}_{s}\mathbf{v})\ \mathbf{v} \ \tau \Big)
+\hbox{ div}\Big( \rho (\mathbf{v}_{s}\mathbf{v})\ \mathbf{v}\Big)
\tau
\]
\[
-{\frac{\partial }{\partial t}}\Big( \rho
_{n}(\mathbf{v}_{n}-\mathbf{v} _{s})\mathbf{v}_{n}\tau \Big)
+{\frac{\partial }{\partial t}}\Big( \rho
_{n}(\mathbf{v}_{n}-\mathbf{v}_{s})\mathbf{v}_{n}\Big) \tau
-\hbox{ div} \Big( \rho
_{n}\mathbf{v}_{n}(\mathbf{v}_{n}-\mathbf{v}_{s})\mathbf{v}
_{n}\tau \Big)
\]
\[
+\hbox{ div}\Bigg( \rho
_{n}\Big((\mathbf{v}_{n}-\mathbf{v}_{s})\mathbf{v} _{n}\Big)
\mathbf{v}_{n}\Bigg) \tau +\hbox{ div}\left( \rho R \mathbf{v}\
\tau \right) -\hbox{ div}\left( \rho R\ \mathbf{v}\right) \tau
\]
\[
+\hbox{ div}\left( \rho _{n}R_{n}\mathbf{v}_{n}\tau \right)
-\hbox{ div }\left( \rho _{n}R_{n}\mathbf{v}_{n}\right) \tau
+{\frac{\partial }{\partial t}}\left( L \tau \right)
-{\frac{\partial L}{\partial t}}\tau.
\]
Consequently,
\[
{\frac{\partial }{\partial t}}\left( \rho
\mathbf{v}_{s}\mathbf{v}+\rho _{n}(
\mathbf{v}_{n}-\mathbf{v}_{s})\mathbf{v}_{n}-L\right) +
\]
\[
\hbox{ div}\left\{ \left( \mathbf{v}_{s}\mathbf{v}-R\right) \rho
\mathbf{v} +\left[
(\mathbf{v}_{n}-\mathbf{v}_{s})\mathbf{v}_{n}-R_{n}\right] \rho
_{n} \mathbf{v}_{n}\right\} =0
\]
If we notice that
\[
\rho \mathbf{v}_{s}\mathbf{v}+\rho
_{n}(\mathbf{v}_{n}-\mathbf{v}_{s})
\mathbf{v}_{n}-L={\frac{1}{2}}\rho _{n}v_{n}^{2}+{\frac{1}{2}}\rho
_{s}v_{s}^{2}+ \beta =\rho \varepsilon
\]
and
\[
(\mathbf{v}_{s}\mathbf{v}-R)\rho \mathbf{v}+\Big(
(\mathbf{v}_{n}-\mathbf{v} _{s})\mathbf{v}_{n}-R_{n}\Big) \rho
_{n}\mathbf{v}_{n}=
\]
\[
({\frac{1}{2}}v_{s}^{2}+\beta_{\rho _{s}}^{\prime })\rho
_{s}\mathbf{v}_{s}+({ \frac{1}{2}}v_{n}^{2}+\beta_{\rho
_{n}}^{\prime })\rho _{n}\mathbf{v}_{n}= \mathbf{q},
\]
we obtain the equation of balance of the total energy in the form
: \begin{equation} {\frac{{\partial \rho \varepsilon }}{\partial
t}}+\hbox{ div}\ \mathbf{q}=0. \label{heat}
 \end{equation}
 We
notice that the specific entropy $S$ does not appear explicitly
anymore in equations (\ref{irrot1}), (\ref{momentum3}),
(\ref{heat}) and we conclude : \emph{In the case of superfluid
Helium the }\textsc{Hamilton}\emph{\ principle yields} the
\textsc{Landau} \emph{model.}

\section{Appendix. Variation of Basic Tensorial Quantities}

Let \ $\left( \lambda ,\mathbf{X}\right) \ $ be any generalized
 Lagrangian coordinates and $\left( t,\mathbf{x}\right) \ $ the
associated Eulerian coordinates
\begin{equation}
\cases{t=g\left( \lambda ,\mathbf{X}\right) \cr
\mathbf{x}={\mbox{\boldmath$\phi$}}\left( \lambda
,\mathbf{X}\right) . \label{mov}}
\end{equation} The relation $
d\mathbf{x}=\mathbf{v}\ dt+Fd\mathbf{X} $ defines simultaneously
the velocity vector and the deformation gradient of motion
(\ref{mov}) :
\[
\mathbf{v}={\frac{\partial {\mbox{\boldmath$\phi$}}}{\partial
\lambda }}{ \frac{1}{\displaystyle {\frac{\partial g}{\partial
\lambda }}}}, \quad \mathbf{F}={\frac{\partial
{\mbox{\boldmath$\phi$}}}{\partial \mathbf{X}}}-{ \frac{\partial
{\mbox{\boldmath$\phi$}}}{\partial \lambda }}{\frac{\partial g
}{\partial \mathbf{X}}}{\frac{1}{\displaystyle {\frac{\partial
g}{\partial \lambda }}}}.
\]
Let \ $\displaystyle \cases{t=G\left( \lambda
,\mathbf{X},\varepsilon \right) \cr
\mathbf{x}={\mbox{\boldmath$\Phi$}}\left( \lambda
,\mathbf{X},\varepsilon \right) }\ $\ be a virtual motion. The
associated perturbation of the velocity   $\mathbf{v} $ is given
by the formula :
\[
\mathbf{u}={\frac{\partial {\mbox{\boldmath$\Phi$}}}{\partial
\lambda }}{ \frac{1}{\displaystyle {\frac{\partial G}{\partial
\lambda }}}}
\]
and consequently,
\[
\delta \mathbf{v}=\left. {\frac{d\mathbf{u}}{d\varepsilon
}}\right| _{\varepsilon =0}={\frac{\partial
{\mbox{\boldmath$\xi$}}}{\partial \lambda }
}{\frac{1}{\displaystyle {\frac{\partial g}{\partial \lambda
}}}}-\mathbf{v}{ \frac{\partial \tau }{\partial \lambda
}}{\frac{1}{\displaystyle  {\frac{
\partial g}{\partial \lambda }}}}.
\]
For fixed values of Lagrangian coordinates the variation of
$\mathbf{v} $ in Eulerian coordinates is :
\[
\delta
\mathbf{v}={\frac{d{\mbox{\boldmath$\xi$}}}{dt}}-\mathbf{v}{\frac{
d\tau }{dt}}\hskip 1cm\hbox{where}\hskip
1cm{\frac{d}{dt}}={\frac{\partial }{
\partial t}}+\mathbf{v}{\frac{\partial }{\partial \mathbf{x}}}.
\]
Analogous calculation for \ $\mathbf{F}$ is :
\[
\delta \mathbf{F}=\left( {\frac{\partial
{\mbox{\boldmath$\xi$}}}{\partial
\mathbf{x}}}-\mathbf{v}{\frac{\partial \tau }{\partial
\mathbf{x}}}\right) \mathbf{F}.
\]
Moreover, the \textsc{Euler}-\textsc{Jacobi} identity yields
\[
\delta \hbox{ det }\mathbf{F}=\hbox{ det }\mathbf{F}\ \hbox{ tr
}\left( \mathbf{F}^{-1}\delta \mathbf{F}\right).
\]
Hence, the mass conservation law is :
$ \rho \hbox{ det
}\mathbf{F}=\rho _{0}\left( \mathbf{X}\right) $ and implies
\[
\delta \rho =-\rho \left( \hbox{ div}\
{\mbox{\boldmath$\xi$}}-\nabla \tau \cdot \mathbf{v}\right)
\]
Equation (\ref{table1})$_2$ is the form of the mass balance for
the normal component  of Helium. If we assume
\[
\rho _{n}\hbox{ det }\mathbf{F}_{n}=\rho _{0n}\left( \lambda
_{n},\mathbf{ X_{n}}\right),
\]
which means that  $\rho _{n}$ is defined on the Lagrangian space
of the normal component, the variation of $\rho _{n}$ with respect
to $ \delta _{n}$ is always in the form :
\[
\delta _{n}\rho _{n}=-\rho _{n}\left( \hbox{ div}\
{\mbox{\boldmath$\xi$}} _{n}-\nabla \tau \cdot
\mathbf{v}_{n}\right).
\]

\bigskip

{\bf Acknowledgment}: The present paper was developed during the
stay of Henri Gouin as visiting professor in C.I.R.A.M. of the
University of Bologna with a fellowship of the Italian C.N.R.


\begin{thebibliography}{99}
\bibitem{trusd}  {\normalsize C. Truesdell, \emph{Sulle basi della
termomeccanica}, Rend. Accad. Naz. Lincei  \textbf{8}, 158 (1957).
}

\bibitem{ingomixture}  {\normalsize I. M\"{u}ller, \emph{A new approach to
thermodynamics of simple mixtures}, Zeitschrift f\"{u}r
Naturforschung \textbf{28a}, 1801 (1973). }

\bibitem{koli}  {\normalsize K. Hutter, I. M\"{u}ller, \emph{On mixtures of
relativistic fluids}, Helvetica physica Acta  \textbf{48}, 675
(1975). }

\bibitem{dreyer}  {\normalsize W. Dreyer, \emph{Zur Thermodynamik von Helium
II - Superfluides Helium mit und ohne Wirbellinien als bin\"{a}re
Mischung}, Dissertation Technische Universit\"{a}t Berlin (1983).
}

\bibitem{45}  {\normalsize {I. M\"{u}ller,} \emph{Thermodynamics}, Pitman,
New York (1985). }

\bibitem{landau}  {\normalsize {L. Landau, }{E. Lifchsitz}, \textit{M\'{e}canique des Fluides}, p. 418, Mir, Moscow (1971). }

\bibitem{putterman}  {\normalsize S. Putterman, \emph{Super Fluid
Hydrodynamics}, Elsevier, New York (1974).}

\bibitem{rogers}  {\normalsize {T. Ruggeri,} \emph{The binary mixtures of
Euler fluids: A unified theory of second sound phenomena} in
Continuum Mechanics and Applications in Geophysics and the
Environment, Eds. B. Straughan, R. Greve, H. Ehrentraut and Y.
Wang, p. 79, Springer-Verlag, Berlin (2001).}

\bibitem{ggp}  {\normalsize {S. Gavrilyuk}, {H. Gouin}, {Yu. Perepechko},
\emph{Hyperbolic models of homogeneous two-fluid mixtures},
Meccanica \textbf{33}, 161 (1998). }

\bibitem{gg1}  {\normalsize {H. Gouin}, {S. Gavrilyuk}, \emph{Hamilton's
principle and Rankine-Hugoniot conditions for general motions of
mixtures}, Meccanica  \textbf{34}, 39 (1999).}

\bibitem{gg2}  {\normalsize {S. Gavrilyuk}, {H. Gouin}, \emph{A new form of
governing equations of fluids arising from Hamilton's principle},
Int. J. Eng. Sci.  \textbf{37}, 1495 (1999). }

\bibitem{g}  {\normalsize {H. Gouin}, \emph{Variational theory of mixtures
in continuum mechanics}, Eur. J. Mech. B/Fluids  \textbf{9}, 469
(1990). }

\bibitem{et}  {\normalsize {I. M\"{u}ller}, {T. Ruggeri}, \emph{Rational
Extended Thermodynamics}, 2nd ed., Springer Tracts in Natural
Philosophy \textbf{37}, Springer-Verlag, New York (1998). }

\bibitem{rugg4}  {\normalsize {T. Ruggeri,} {A. Muracchini}, {L. Seccia},
\emph{A Continuum Approach to Phonon Gas and Shape Changes of
Second Sound via Shock Waves Theory}, Nuovo Cimento D \textbf{16},
15 (1994). }

\bibitem{rugg3}  {\normalsize {T. Ruggeri}, {A. Muracchini}, {L. Seccia},
\emph{Shock Waves and Second Sound in a Rigid Heat Conductor: A
Critical Temperature for NaF and Bi}, Phys. Rev. Lett.
\textbf{64}, 2640 (1990). }

\bibitem{rugg5}  {\normalsize {T. Ruggeri, } {A. Muracchini}, {L. Seccia},
\emph{Second Sound and Characteristic Temperature in Solids},
Phys. Rev. B \textbf{54}, 332 (1996). }

\bibitem{rugg7}  {\normalsize {T. Ruggeri,} {A. Muracchini}, {L. Seccia},
\emph{Second Sound Propagation in Superfluid Helium via Extended
Thermodynamics, }}Lecture notes{\normalsize \emph{\ }WASCOM 2001.}

\bibitem{serrin}  {\normalsize {J. Serrin,} \emph{Mathematical principles of
classical fluid mechanics},   Encyclopedia of Physics, vol.
VIII/I, p. 125-263, Springer-Verlag, Berlin (1959).}
\end{thebibliography}
\end{document}